\documentclass[12pt]{article}
\usepackage[utf8]{inputenc}
\usepackage[T1]{fontenc}
\usepackage{geometry}
\usepackage{newtxtext}
\usepackage{newtxmath}
\usepackage{amsmath}
\usepackage{graphicx}
\usepackage{titlesec}
\usepackage{setspace}
\usepackage{bm}
\usepackage{hyperref}
\hypersetup{
    colorlinks=false
}
\usepackage[english]{babel}

\titleformat{\section}[runin]
  {\normalfont\normalsize\itshape}{\indent\normalfont\thesection.}{0.5em}{}[]
\titleformat{\subsection}[runin]
  {\normalfont\normalsize\itshape}{\indent\normalfont\thesubsection.}{0.5em}{}
\titlespacing*{\section}{0pt}{0.0\baselineskip}{0.35\baselineskip}
\titlespacing*{\subsection}{0pt}{0.0\baselineskip}{0.35\baselineskip}

\newcommand{\de}{\mathrm{d}}
\newcommand{\phin}{\varphi_\textsc{N}}
\newcommand{\bppn}{\beta_\textsc{ppn}}
\newcommand{\gppn}{\gamma_\textsc{ppn}}
\newcommand{\kms}{\textrm{km}\,\textrm{s}^{-1}}

\title{PPN MOTION OF THE S-STARS AROUND SGR A*}
\author{R. I. GAINUTDINOV$^{1,2}$} 
\date{Received ???\\ Accepted ???}

\begin{document}

{\setstretch{1.0}
\maketitle
\setstretch{1.0}
\small \indent 
Parametrized  Post-Newtonian (PPN) equations of motion for several S-stars nearest to  the Galactic Center 
 compact relativistic object SgrA* are considered. 
The effect of the orbital periods difference between Newtonian and Post-Newtonian cases is taken into account. The best fit PN orbit of S2 has a period which is 16 days longer than Newtonian one. The PPN parameters $\bppn$ and $\gppn$ are measured. Bayesian sampling is used to fit the trajectories of the PPN laws of motion. Posterior estimates of $\bppn$ and $\gppn$ are $0.97^{+0.42}_{-0.65}$ and $0.81^{+0.46}_{-0.60}$ respectively. The result confirms General Relativity prediction for the Post-Newtonian equations of motion in the conditions of orbital motions in vicinity of the SgrA*.\par
}
\normalsize
\vspace{0.6cm}
\section{Introduction.} The S-stars is a name given to a cluster of stars orbiting the supermassive relativistic compact object Sgr A* at the Galactic Center. For the most part orbits of these stars have high eccentricities. So they reach high precientral velocities ($\sim$ 0.01 the speed of light). This fact along with close location to supermassive object makes the S-star cluster a unique observational object to investigate. The values of $v^2/c^2$ and $\phin/c^2$ are $\sim 10^{-4}$, while these parameters for binary neutron star orbits have values about $\sim 10^{-6}$ . Thus the Post-Newtonian S-stars orbital corrections test new region of relativistic gravity effects. We consider implications of the Post-Newtonian laws of orbital motion and what kind of visible effects make the PN motion different from Newtonian. Post-Newtonian effects can also be used to check different gravitational theories. In this paper we obtain the parameters of Parametrized Post-Newtonian formalism: $\bppn$ and $\gppn$. Though these parameters were measured in solar system and binary neutron stars, now we can estimate them from direct measurements of the PN orbit parameters without pulsar timing and fitting timing model. Modern observations of the S-stars \cite{Boehle}, \cite{Chu}, \cite{Do}, \cite{Gillessen2017} provide us with this opportunity.\par
The observations have started in 1992. Since then the time series arrays of observational data became thorough enough. The S-stars cluster was often used in different investigations: \cite{Eckart}, \cite{Genzel}, \cite{Gillessen2009a}, \cite{Gillessen2009b}, \cite{Habibi}, \cite{Lacroix}, \cite{Mouawad}, \cite{Schodel2003a}, \cite{Schodel2003b}. In the work \cite{ZT} the authors have shown that it is unlikely to detect a star that is closer to gravitation center than any of other observed S-stars.\par
The star S2 (also known as S0-2) is the most popular of S-stars as it has accurately observed positions and radial velocities and it also has one of the shortest orbital periods: $\sim$ 16 years. It passed own pericenter twice since the beginning of the observations. The first pericenter passage was at 2002. The second one was at May, 2018. The S2 star played a key role in many studies: the measurement of the distance to the Galactic Center $R_0$ and the mass of Sgr A* $\mathfrak{M}$ \cite{Eisenhauer}, \cite{Gillessen2017}, \cite{GRAVITY2019a}; investigating the spin of the Galactic Center Black Hole \cite{Iorio2011b}, \cite{JBH}, \cite{Waisberg}, \cite{ZLY}; investigating accretion flow and stellar winds \cite{Christie}, \cite{SBB}; applying different gravitation theories: \cite{Borka2013}, \cite{Zakharov2018a}, \cite{Zakharov2018b} -- Yukawa gravity, \cite{GRAVITY2019b} -- scalar field, \cite{Borka2012}, \cite{Zakharov2014} -- $R^n$ gravity, \cite{Dialektopoulos} -- non-local gravity, and other gravitation theories \cite{Borka2016}, \cite{RS}, \cite{Zakharov2016}; testing GR effects \cite{Grould}, \cite{Iorio2011a}; and other investigations \cite{GGM}, \cite{Lacroix}.\par
The stars S2 (also known as S0-2), S38 and S55 (also known as S0-102) are known for their small orbital periods. Thus they are the stars of our greatest interest. The S2 pericenter passage in May, 2018 was used to make gravitational redshift research \cite{Do}, \cite{GRAVITY2018}. The star S55 has the smallest orbital period of $\sim$ 12 years.\par
Close distances to supermassive black hole and high velocities of S-stars provide a basis for Post-Newtonian motion research \cite{ZI}, \cite{Parsa}, \cite{PS}, \cite{Saida}, which is also a topic of our work. We consider not only Post-Newtonian but also the Parametrized Post-Newtonian laws of motion. Bayesian techniques are used to obtain $\beta_\textsc{PPN}$ and $\gamma_\textsc{PPN}$ estimates by the posterior distributions.\par

\section{Observational data.} The S-stars observations consist of two types of data: visual position of the stars -- astrometric data (RA $\alpha$ and Dec $\delta$ difference from RA and Dec of Sgr A*), and the radial velocities (RV) of the stars -- spectroscopic data. The Sgr A* is also moving on the plane of sky. So the visual positions of the stars are relative to the visual position of Sgr A* for the initial epoch (1992.224). Radial velocities are also measured with the consideration of non-relativistic Doppler effect. So we cannot just interpret modelled radial velocities as observed radial velocities. We must transform them in the way which is presented by the formula (\ref{RVmodel}).\par
Modern observations are presented in \cite{Gillessen2017}, \cite{Boehle}, \cite{Chu}, \cite{Do}. The data arrays presented in the work \cite{Gillessen2017} are the most thorough. From there we will take the data sets of S2 (145 astrometric and 44 RV measurements), S38 (116 astrometric and 5 RV measurements), and S55 (44 astrometric and 2 RV measurements). The arrays are obtained using VLT. The work \cite{Chu} contains 12 new RV measurements for S2, while other measurements are taken from \cite{Boehle} and \cite{Gillessen2017}. \cite{Boehle} contains 34 RV measurements for S2, 1 RV for S38, and 34 astrometric measurements for S38. These data sets are obtained by W. M. Keck observatory. The work \cite{Do} contains the last data of S2 star. 11 astrometric and 28 RV measurements are new. Some of the RV measurements of this work are obtained with the Subaru telescope.\par
Totally we have 156 positions and 118 RV's for S2, 150 positions and 6 RV's for S38, and 44 positions and 2 RV's for S55. All of RV measurements are VLSR-corrected. The sets of astrometric data from the different papers contain a small error which is caused by the uncertainties in estimating the location of central mass. So the arrays of visual positions from different works are presented in slightly different reference frames.\par

\section{Orbital fitting.}
\subsection{Methods of comparison.} The Keplerian orbit in 2 dimensions is defined by 4 parameters: major semi-axis $a$, eccentricity $e$, pericenter argument $\omega$, and an epoch of pericenter passage $t_\text{per}$. These parameters can be transformed to the phase space vector $(x, y, \dot{x}, \dot{y})$ for a fixed epoch. Keplerian parameters are constant if we use Newtonian laws of motion, but in Post-Newtonian case they become osculating. Thus it makes sense for us to use another 4 parameters instead. They are the components of the initial phase space $(x_0, y_0, \dot{x}_0, \dot{y}_0)$. We will use these parameters as an initial condition for the 4 order Runge-Kutta integrator, which will produce a modelled trajectory in the own plane. And to rotate the plane itself, we have the remaining 2 Keplerian parameters: inclination $i$ and a longitude of the ascending node $\Omega$.\par
Using the same initial conditions $(x_0, y_0, \dot{x}_0, \dot{y}_0)$ must lead to the different resulting trajectories for Newtonian and Post-Newtonian cases. The Post-Newtonian orbit appears to have a bigger apocenter. It also has a pericenter shift effect, but it is small.\par
\begin{table}[ht]
\label{tab:1}
\caption{\emph{S-stars pericenter shift values}}
\vspace{0.1cm}
\centering
\begin{tabular}{cccc}
\hline
Star & S2  & S38 & S55 \\ \hline
$\Delta\omega$ & $12'$ & $7.1'$ & $6.7'$ \\
$\dot{\omega}$ & $45''/\mathrm{yr}$ & $22''/\mathrm{yr}$ & $31''/\mathrm{yr}$ \\
$\dot{\omega}\cdot 100\,\mathrm{yrs}$ & $1^\circ$ 15' & $37'$ & $52'$ \\
\hline
\end{tabular}
\end{table}
Table 1 presents some of the S-stars pericenter shift values. The $\Delta\omega$ value is a pericenter shift per one synodic period. These values are barely detectable even after 100 years of observations.\par
We can also use different initial conditions $(x_0, y_0, \dot{x}_0, \dot{y}_0)$ for Newtonian and Post-Newtonian laws of motion to get the orbits of the same size. The difference is that the PN pericentral velocity would be less than the Newtonian. So these trajectories would have different orbital periods. The Post-Newtonian period has to be bigger. So we have a purely kinematic effect of some kind of a `delay'. It is a very significant qualitative difference between the Newtonian and Post-Newtonian motion. In our case, we will use the MCMC sampler to obtain the best fit Newtonian and Post-Newtonian trajectory to compare their orbital periods.
\par
\subsection{PPN Equations of motion.} We want to take a look at the Schwarzschild solution of Einstein's field equations
\begin{equation}\label{Schwsol}
    \de s^2 = \Big( 1 - \frac{2\mu}{r} \Big) c^2 \de t^2 - \frac{\de r^2}{\big( 1 - \frac{2\mu}{r} \big)} - r^2 \de \Omega^2,
\end{equation}
where $\mu = GM / c^2$ is a \emph{gravitational radius} and $\de \Omega^2 = \de \theta^2 + \sin^2 \theta \de \phi^2$ is the metric on a sphere of a unit radii. The coordinate system used here is called standard (or Schwarzschild) coordinate system $(t, r, \theta, \phi)$. Angular coordinates $\theta, \phi$ have their usual geometric sense of polar angle and azimuthal angle. The circle of equal radial coordinate $r$ has a length of $2\pi r$. But $r$ is not equal to the distance from the circle to its center. This is the geometric sense of the Schwarzschild coordinates \cite{LL}.\par

We shall use the isotropic coordinate system $(t, \rho, \theta, \phi)$. It is commonly used in relativistic celestial mechanics problems, while Schwarzschild coordinates are considered to be "too accurate" (\cite{MTW}, p.1097). The Schwarzschild-to-isotropic coordinate transformation affects only the radial coordinate
\begin{equation}\label{r-to-rho}
    r=\rho\Big( 1 + \frac{\mu}{2\rho} \Big)^2,
\end{equation}
so $r$ and $\rho$ are asymptotically equivalent. However the directly observed distances between SgrA* and the S-stars are taken from ordinary Euclidean  geometry in our remote observer system. Hence the evolution of orbital parameters are subjected to the specific equations of motion. 

The advantage of the isotropic coordinate system can be seen in the metric
\begin{equation}\label{Schwiso}
    \de s^2 = \frac{\big( 1 - \frac{\mu}{2\rho} \big)^2}{\big( 1 + \frac{\mu}{2\rho} \big)^2}c^2 \de t^2 - \Big( 1 + \frac{\mu}{2\rho} \Big)^4(\de \rho^2 + \rho^2 \de \Omega^2).
\end{equation}
The spatial part of this metric is conformally Euclidean. Thus we can transform spherical isotropic coordinates $(t, \rho, \theta, \phi)$ to Cartesian isotropic coordinates $(t, x, y, z)$ without losing the form of the metric. These coordinates are different from the Cartesian coordinates in their usual sense. But in our approximation  we can consider them as usual Cartesian coordinates. The metric in this system is
\begin{equation}\label{SchwisoCart}
    \de s^2 = \frac{\big( 1 - \frac{\mu}{2|\mathbf{x}|} \big)^2}{\big( 1 + \frac{\mu}{2|\mathbf{x}|} \big)^2}c^2 \de t^2 - \Big( 1 + \frac{\mu}{2|\mathbf{x}|} \Big)^4\de \mathbf{x}^2.
\end{equation}
\indent Next we want to define a value of the Newtonian potential $\phin = -GM / \rho = -c^2 \mu / |\mathbf{x}|^2$
\begin{equation}\label{SchwisoCart2}
    \de s^2 = \frac{\big( 1 + \frac{\phin}{2c^2} \big)^2}{\big( 1 - \frac{\phin}{2c^2} \big)^2}c^2 \de t^2 - \Big( 1 - \frac{\phin}{2c^2} \Big)^4\de \mathbf{x}^2.
\end{equation}
\indent To perform the first order Post-Newtonian expansion one should expand $g^{00}$ to the order of $O(c^{-6})$ and $g^{xx}$ to the order of $O(c^{-4})$
\begin{equation}\label{SchwPN}
    \de s^2 = \bigg( 1 + \frac{2\phin}{c^2} + \frac{2\phin^2}{c^4} + O(c^{-6}) \bigg) c^2 \de t^2 - \bigg( 1 - \frac{2\phin}{c^2} + O(c^{-4}) \bigg) \de \mathbf{x}^2.
\end{equation}
\indent In the frame of Parametrized Post-Newtonian formalism this expansion looks like
\begin{equation}\label{SchwPPN}
    \de s^2 = \bigg( 1 + \frac{2\phin}{c^2} + \bppn\frac{2\phin^2}{c^4} + O(c^{-6}) \bigg) c^2 \de t^2 - \bigg( 1 - \gppn\frac{2\phin}{c^2} + O(c^{-4}) \bigg) \de \mathbf{x}^2,
\end{equation}
where the values of $\bppn$ and $\gppn$ are different for different gravitation theories. The case of $\bppn=1$ and $\gppn=1$ is consistent with General Relativity.\par
Dividing (\ref{SchwPPN}) by $c^2 \de t^2$ implies
\begin{equation}\label{SchwPPN2}
\begin{aligned}
    \frac{1}{c^2}\Big(\frac{\de s}{\de t}\Big)^2 &= \bigg( 1 + \frac{2\phin}{c^2} + \bppn\frac{2\phin^2}{c^4} + O(c^{-6}) \bigg) - \bigg( 1 - \gppn\frac{2\phin}{c^2} + O(c^{-4}) \bigg) \frac{\dot{\mathbf{x}}^2}{c^2}=\\
    &=1 - \frac{\dot{\mathbf{x}}^2}{c^2} + \frac{2\phin}{c^2} + \gppn \frac{2\phin \dot{\mathbf{x}}^2}{c^4} + \bppn \frac{2\phin}{c^4} + O(c^{-6}).
\end{aligned}
\end{equation}
\indent Taking the square root with the accuracy of $O(c^{-6})$ leads to
\begin{equation}\label{dsdt}
    \frac{1}{c}\frac{\de s}{\de t} = 1 - \frac{\dot{\mathbf{x}}^2}{2c^2} + \frac{\phin}{c^2} - \frac{\dot{\mathbf{x}}^2}{8c^4} + (1 + 2\gppn)\frac{\phin\dot{\mathbf{x}}^2}{2c^4} + (2\bppn - 1)\frac{\phin^2}{2c^4} + O(c^{-6}).
\end{equation}
The equations of motion can be derived from the variational principle
\begin{equation}\label{varpr}
    \delta \int \de s = \delta \int \Big( \frac{\de s}{\de t} \Big) \de t = 0,
\end{equation}
where $\de s/ \de t$ is given by the equation (\ref{dsdt}). Multiplying (\ref{dsdt}) by $-c^2$ and getting rid of a constant term and $O(c^{-6})$ produces a Lagrangian
\begin{equation}\label{Lagrangian}
    L = \frac{\dot{\mathbf{x}}^2}{2}\bigg( 1 + \frac{\dot{\mathbf{x}}^2}{4c^2} - (1 + 2\gppn)\frac{\phin}{c^2} \bigg) - \phin\bigg(1 + (2\bppn - 1)\frac{\phin}{2c^2} \bigg).
\end{equation}
\indent Corresponding equations of motion are
\begin{equation}\label{eqmot}
    \ddot{\mathbf{x}} = - \bm{\nabla}\phin \bigg( 1 + 2(\bppn + \gppn) \frac{\phin}{c^2} + \gppn \frac{\dot{\mathbf{x}}^2}{c^2} \bigg) + (2\gppn + 2)\bigg( \bm{\nabla}\phin \cdot \frac{\dot{\mathbf{x}}}{c} \bigg)\frac{\dot{\mathbf{x}}}{c}.
\end{equation}
\par
These are the PPN equations. To obtain the PN equations one must substitute values $\bppn=1$ and $\gppn=1$
\begin{equation}\label{eqmotPN}
    \ddot{\mathbf{x}} = - \bm{\nabla}\phin \bigg( 1 + 4 \frac{\phin}{c^2} + \frac{\dot{\mathbf{x}}^2}{c^2} \bigg) + 4\bigg( \bm{\nabla}\phin \cdot \frac{\dot{\mathbf{x}}}{c} \bigg)\frac{\dot{\mathbf{x}}}{c}.
\end{equation}
\par
Both of these equations (\ref{eqmot}), (\ref{eqmotPN}) give us the Newtonian equations of motion $(\ddot{\mathbf{x}}=-\bm{\nabla}\phin)$ if we consider a limit of infinite speed of light.

\subsection{Light propagation.} It is incorrect to interpret simulated trajectory as an observable one, because of Post-Newtonian effects related to light propagation.\par
S-stars reach high velocities during the time they pass own pericenters. It means that the \emph{Relativistic Doppler effect} should be taken into account. The frequency change formula is given by
\begin{equation}\label{Dopp}
    \omega_0 = \omega\frac{1-\frac{v}{c}\cos \alpha}{\sqrt{1-\frac{v^2}{c^2}}} = \omega\frac{1+\mathrm{RV}_\text{model}/c}{\sqrt{1-\frac{v^2}{c^2}}} \approx \omega \bigg( 1 + \frac{\mathrm{RV}_\text{model}}{c} + \frac{v^2}{2c^2}   \bigg),
\end{equation}
where $\omega$ is a received frequency, $\omega_0$ is an emitted frequency, $\vec{v}$ is a velocity of the source relative to the observer, and $\alpha$ is an angle between the velocity vector $\vec{v}$ and the direction from source to the observer.\par
S-stars are located close to supermassive object. The gravitational field there is strong enough for \emph{gravitational redshift} to become an observable effect. So we must take it into account.
\begin{equation}\label{gravsh}
    \omega_0 = \omega \sqrt{g^{00}} \approx \omega\bigg( 1 + \frac{\phin}{c^2}\bigg).
\end{equation}
\indent To sum up, taking both Doppler effect (\ref{Dopp}) and gravitational redshift (\ref{gravsh}) leads to
\begin{equation}\label{RVmodel}
    \mathrm{RV}_\text{obs}/c = \mathrm{RV}_\text{model}/c + \frac{\phin}{c^2} + \frac{v^2}{2c^2}.
\end{equation}
\par
Another light propagation effect is the R\o mer delay, which is caused by the finiteness of the speed of light. The difference between the furthest and the nearest points of an S-star orbit can reach several light days. So this effect must be taken into account.\par
The S-stars are located not so close to Sgr A* for \emph{gravitational lensing} to be strong enough. So we do not take this effect into consideration.\par
\subsection{Brumberg PPN parameters.} The problem of a massless test particle in the static gravitational field was developed by V.A.Brumberg \cite{Brumberg}. He considered the general solution for different coordinate systems. The choice of a specific coordinate system was defined by the $\alpha$ parameter: $\alpha=1$ corresponds to Standard (Schwarzschild) coordinates, $\alpha=0$ corresponds to either harmonic or isotropic coordinates. He then defined another set of parameters ($A,\,B,\,K$) that generalize the problem for another gravitation theories, similar as PPN parameters do. The work \cite{Brumberg} shows the values of these parameters for the General Relativity
\begin{equation}\label{BrumbergPPN-GR}
    A = 2, \quad B=K=2(1-\alpha).
\end{equation}
\par
These parameters depend on the choice of coordinate system $\alpha$. In \cite{Brumberg} the Post-Newtonian equation of motion is derived also for the case of the Schwarzschild coordinates ($\alpha=1$):
\begin{equation}\label{eqmotPN-Sch}
    \ddot{\mathbf{x}} = - \bm{\nabla}\phin \bigg( 1 + 2 \frac{\phin}{c^2} + 2\frac{\dot{\mathbf{x}}^2}{c^2}
    -3\Big(\frac{\mathbf{x}}{x} \cdot
    \frac{\dot{\mathbf{x}}}{c}\Big)
    \bigg) + 2\bigg( \bm{\nabla}\phin \cdot \frac{\dot{\mathbf{x}}}{c} \bigg)\frac{\dot{\mathbf{x}}}{c}.
\end{equation}
In this case the directly observed orbit (by remote observer) can be calculated by integrating 
Eq.(\ref{eqmotPN-Sch}).
Hence the observed evolution of the orbital parameters
for Schwarzschild coordinates ($\alpha=1$) will differ
from the isotropic coordinates ($\alpha=0$), which is conformally Euclidean. 

\par
\subsection{Techniques used.} A model is constructed as follows:
\begin{itemize}
    \item The parameters $(x_0, y_0, \dot{x}_0, \dot{y}_0)$ are used as the initial vector for the 4-order Runge-Kutta integrator. It solves the equations (\ref{eqmotPN}) (or (\ref{eqmot}) for the given parameters of $\bppn, \gppn$) numerically. The central mass of Sgr A* $\mathfrak{M}$ is used as a parameter. It produces an array of vectors $(x, y, \dot{x}, \dot{y})$. This array is a trajectory in its own plane.
    \item The trajectory is rotated with angles $i$ and $\Omega$.
    \item After the rotation, the orbit is transformed to visual observed RA and Dec by dividing by distance to the Galactic Center $R_0$ and adding the proper motion of Sgr A* $\dot{\alpha}_\mathrm{SgrA*}, \dot{\delta}_\mathrm{SgrA*}$.
    \item The integrated radial velocities are transformed to visual measurements of radial velocities $\mathrm{RV}$ with the formula (\ref{RVmodel}). The proper radial velocity of Sgr A* $\dot{r}_\mathrm{SgrA*}$ should be added.
    \item The simulated points of RA, Dec and RV are being spline interpolated. At this step the R\o mer delay is taken into account.
\end{itemize}
As the result, we have a model with 13 parameters. 6 denote the orbit of the star. 2 parameters are $\bppn$ and $\gppn$ that we want to derive. The remaining 5 parameters $(\mathfrak{M}, R_0, \dot{\alpha}_\mathrm{SgrA*}, \dot{\delta}_\mathrm{SgrA*}, \dot{r}_\mathrm{SgrA*})$ are considered as constants. Their values are presented in the work \cite{Gillessen2017}.\par
As we consider 3 stars (S2, S38, S55), we have 18 parameters of their orbits and 2 PPN parameters of $\beta$, $\gamma$. We use the MCMC bayesian sampler (a python package called \emph{emcee}) to find the posterior distributions of our parameters. Thus we will find the estimates.
\section{Results.}
\subsection{Orbital periods difference.} Fig. 1 presents the difference between the best fit Newtonian and Post-Newtonian trajectories of the star S2. The red curve stands for Post-Newtonian case and the black one stands for Newtonian case. We can see that there is an offset between these curves which grows with the time. That is the effect of `delay' which was discussed previously.\par
\begin{figure}[t]
    \centering
    \includegraphics[scale=0.65, clip]{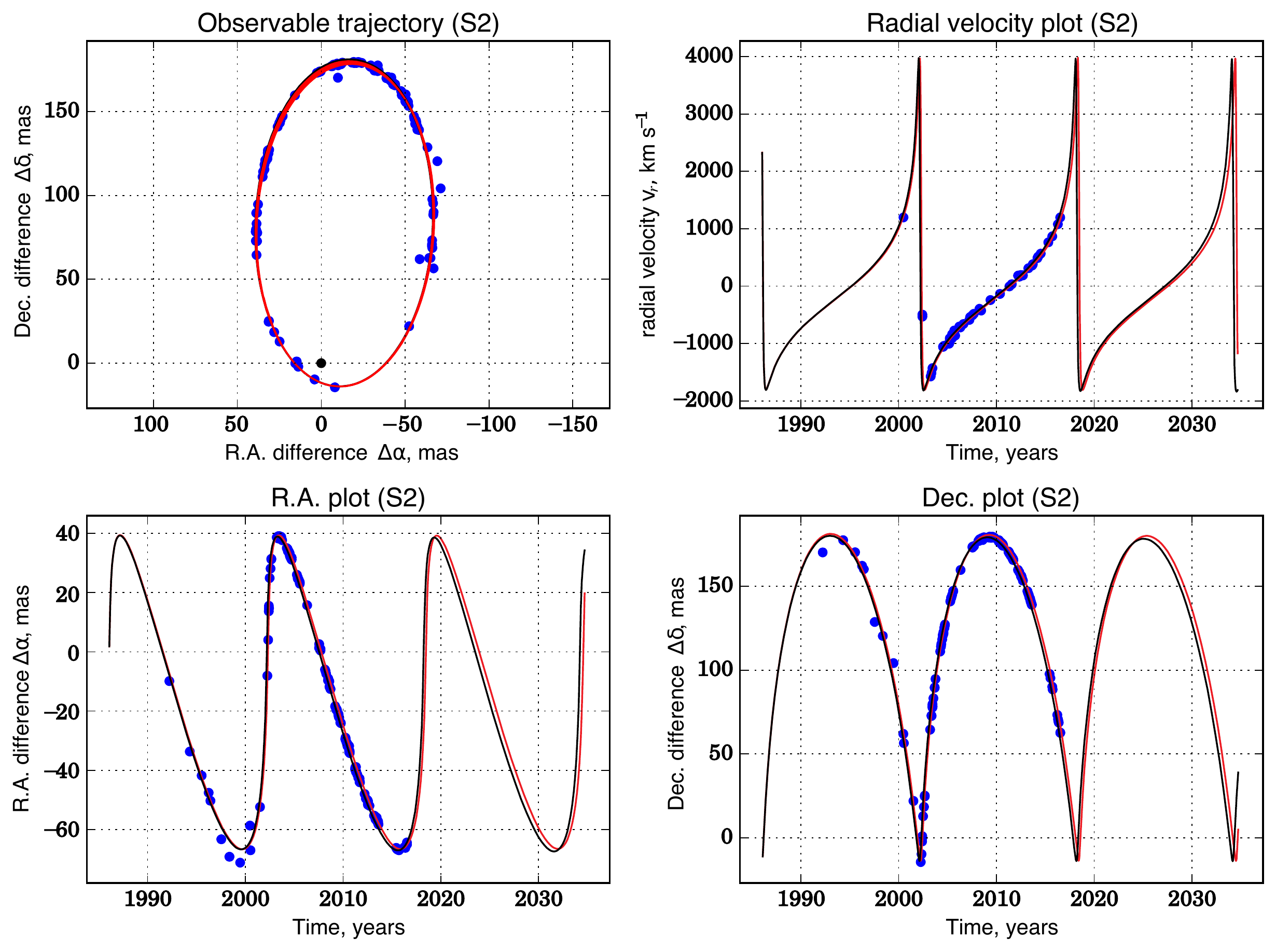}
    \caption{\emph{The difference between the Newtonian and Post-Newtonian motion}}
    \label{fig:1}
\end{figure}\par
The measured value of this difference is 16 days per orbital period. That value is not small and the uncertainties in time axis are very critical when we talk about S-stars. During the pericenter passage the S2 star has a very sharp drop on the RV plot because it moves with a high velocity. The time uncertainty implies misplacing every drop that corresponds to pericenter passage. This effect is even more significant if we talk about future observations, when the S-stars will make several orbital turns.\par
According to our estimates, the date of the next pericenter passage of S2 star is 18 May 2034.
\subsection{PPN parameters estimates.}
We launched the MCMC process that was modelling the motion of S2, S38 and S55 for a given parameters of $\bppn$ and $\gppn$. 
Figure 2 shows the posterior distribution in the $\bppn$ and $\gppn$ parameter space projection.
\begin{figure}[htbp]
    \centering
    \includegraphics[scale=0.8, trim={0.5cm, 0.5cm, 0.5cm, 0cm},clip]{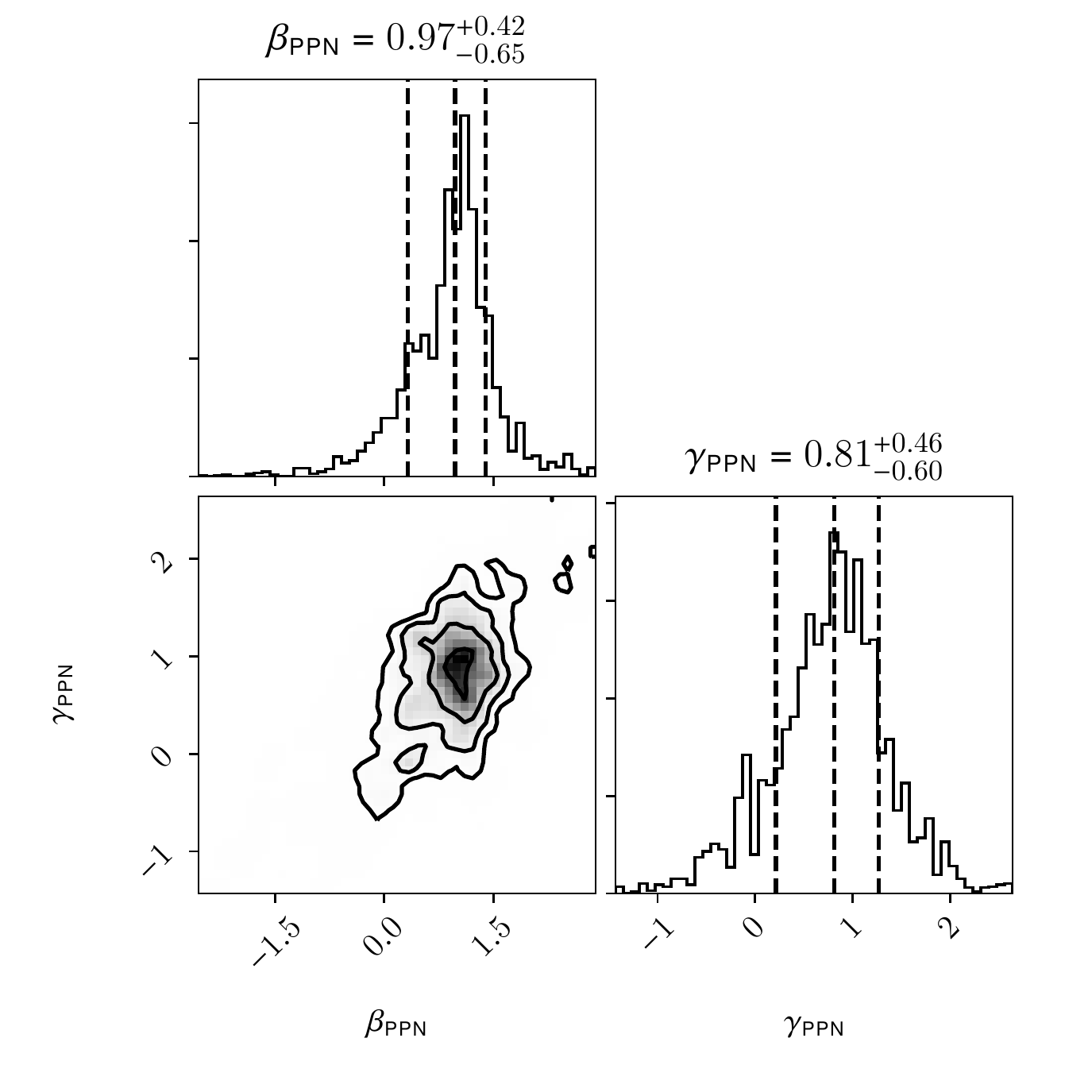}
    \caption{\emph{The posterior distribution. $\bppn, \gppn$ corner plot}}
    \label{fig:2}
\end{figure}\par
The MCMC process was launched with 10000 iterations. Estimated PPN values are
\[\bppn = 0.97^{+0.42}_{-0.65}, \quad \gppn = 0.81^{+0.46}_{-0.60},\]
where the errors are denoted by the 1/6 and 5/6 quantiles of the distribution.\par
Although the result is consistent with General Relativity prediction, this method of obtaining $\bppn$ and $\gppn$ proved to be very inaccurate. The reason is that the PPN parameters deviations do not affect the picture that much.\par
The posterior estimates of S-stars parameters are presented in Table 2.
\begin{table}[ht]
\renewcommand{\arraystretch}{1.3}
\label{tab:2}
\caption{\emph{S-stars parameters}}
\vspace{0.1cm}
\centering
\begin{tabular}{ccccccc}
\hline
Star & $x_0,\;\mu$  & $y_0,\;\mu$ & $\dot{x}_0,\;\kms$ & $\dot{y}_0,\kms$ & $i,\;^\circ$ & $\Omega,\;^\circ$ \\ \hline
S2  & $22954.8^{+4.0}_{-1.5}$ & $37481.5^{+8.9}_{-2.0}$ & $-219.84^{+0.62}_{-0.79}$ & $604.74^{+0.99}_{-0.22}$ & $133.86^{+0.07}_{-0.06}$ & $226.10^{+0.08}_{-0.08}$ \\
S38 & $50102^{+5}_{-2}$ & $69295^{+1}_{-7}$ & $67^{+5}_{-2}$ & $588^{+1}_{-3}$ & $169.9^{+2.6}_{-0.8}$ & $93.39^{+0.14}_{-0.11}$ \\
S55 & $32342^{+8.0}_{-3}$ & $-8724.1^{+0.8}_{-2.1}$ & $-344.8^{+6.8}_{-1.1}$ & $1004.9^{+1.3}_{-2.5}$ & $152.7^{+1.4}_{-1.0}$ & $323.3^{+0.7}_{-0.5}$ \\
\hline
\end{tabular}
\end{table}
We can see that these parameters are defined much more accurately than PPN parameters. Figure 3 shows the best fit trajectories of the three stars and radial velocity plot of S2.

\begin{figure}[htbp]
    \centering
    \includegraphics[scale=0.65, trim={2cm, 0.5cm, 2cm, 0.5cm},clip]{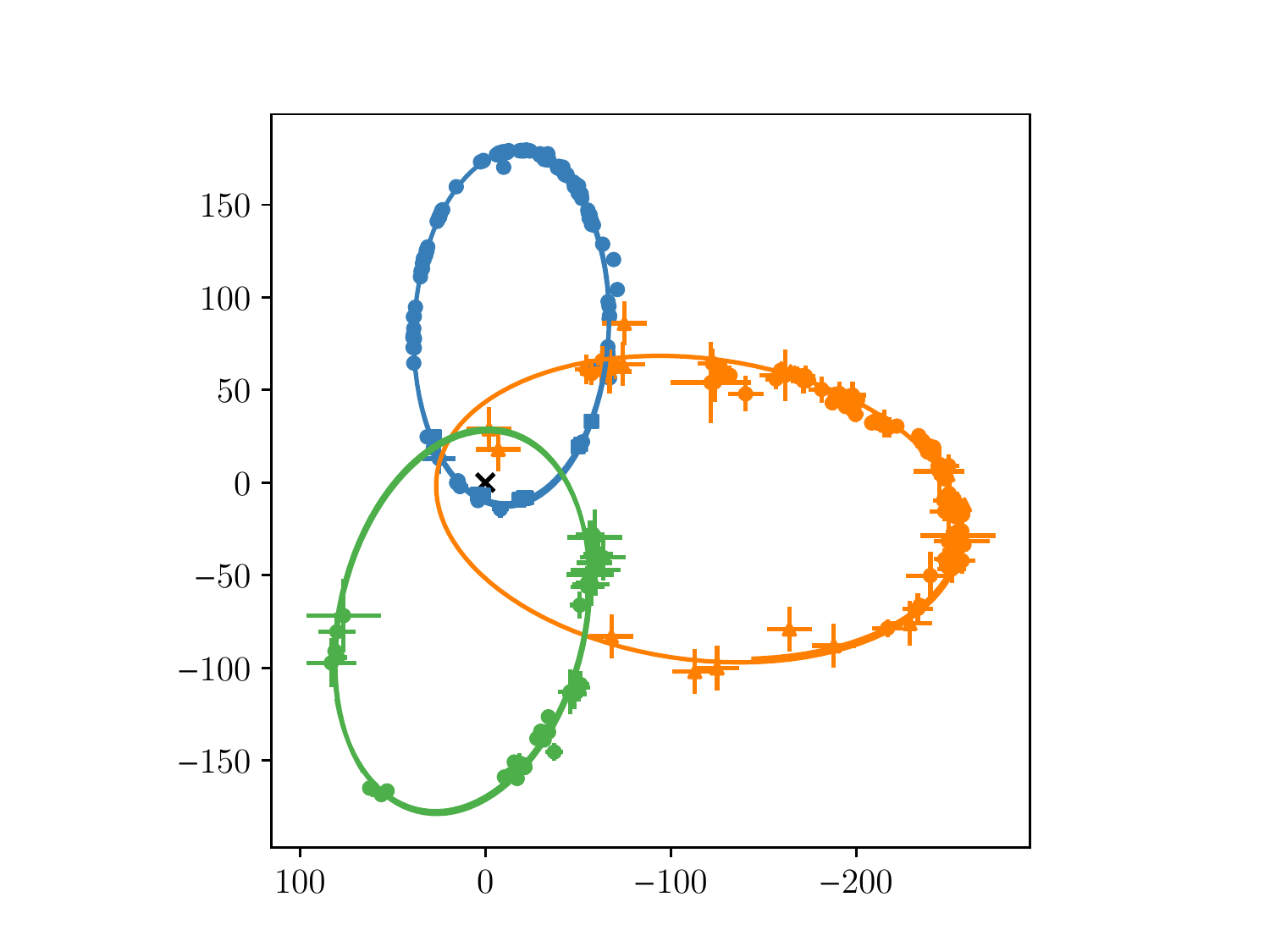}
    \includegraphics[scale=0.65, trim={0.5cm, 0.5cm, 1cm, 0.5cm},clip]{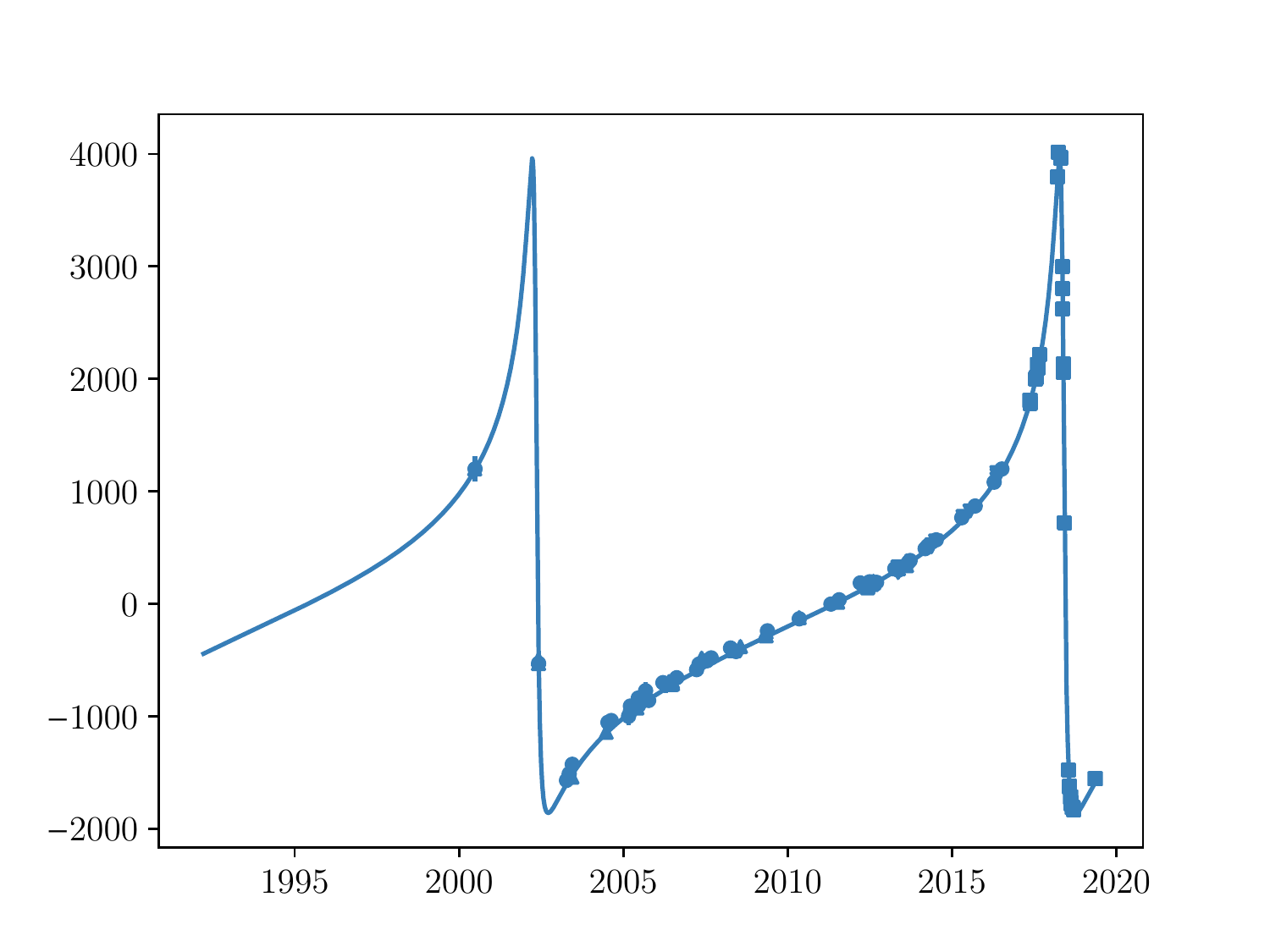}
    \caption{\emph{Left: Trajectories of S2 (blue), S38 (orange) and S55 (green); Right: RV of S2}}
    \label{fig:3}
\end{figure}\par
\section{Conclusion.} In this work we used  the Post-Newtonian equations of motion for analysis of the  orbital parameters of the S-stars.  For S2 star the Post-Newtonian orbital period is 16 days bigger than Newtonian one  (for the same initial conditions). This difference is important because it may produce an uncertainty in the time axis, which is very critical if we talk about the S-stars. Taking this effect into account, we obtained the orbital parameters of S2, S38 and S55. They are presented in the Table 2.  The date of the next pericenter passage of S2 star is 18 May 2034.
\par
We also obtained the estimates of PPN parameters.  Bayesian sampling is used to fit the trajectories of the PPN laws of motion. Posterior estimates of $\bppn$ and $\gppn$ are $0.97^{+0.42}_{-0.65}$ and $0.81^{+0.46}_{-0.60}$ respectively. 
For the S-stars orbital motion the values of $v^2/c^2$ and $\phin/c^2$ are $\sim 10^{-4}$, so our results confirm General Relativity prediction for the Post-Newtonian equations of motion (in isotropic and harmonic coordinates) for the  condition of orbital motions in vicinity of the SgrA*.\par
The work was performed as part of the government contract of the SAO RAS approved by the Ministry of Science and Higher Education of the Russian Federation.

\par
\vspace{0.6cm}
\indent $^{1.}\;$ Special Astrophysical Observatory, Russian Academy of Sciences, Russian Federation,\\
\indent $\;\;\;$ email: \href{mailto:roustique.g@gmail.com}{roustique.g@gmail.com}\\
\indent $^{2.}\;$ Saint Petersburg State University, Russian Federation

\end{document}